\documentclass[twocolumn,showpacs,aps,prl]{revtex4}
\usepackage{graphicx}
\usepackage{dcolumn}
\usepackage{amsmath}
\usepackage{latexsym}

\begin {document}


\title {Information sharing and sorting in a community}
\author
{Biplab Bhattacherjee$^1$, S. S. Manna$^1$ and Animesh Mukherjee$^2$
}
\affiliation
{
\begin {tabular}{c}
$^1$Satyendra Nath Bose National Centre for Basic Sciences,
Block-JD, Sector-III, Salt Lake, Kolkata-700098, India \\
$^2$Department of Computer Science and Engineering, Indian Institute of Technology, Kharagpur-721302, India \\
\end{tabular}
}
\begin{abstract}
      We present the results of detailed numerical study of a model for the sharing and sorting of informations in a
   community consisting of a large number of agents. The information gathering takes place in a sequence of mutual
   bipartite interactions where randomly selected pairs of agents communicate with each other to enhance their 
   knowledge and sort out the common information. Though our model is less restricted compared to the well established 
   naming game, yet the numerical results strongly indicate that the whole set of exponents characterizing this 
   model are different from those of the naming game and they assume non-trivial values. Finally it appears that in analogy
   to the emergence of clusters in the phenomenon of percolation, one can define clusters of agents here having the same information. 
   We have studied in detail the growth of the largest cluster in this article and performed its finite-size scaling analysis.
\end{abstract}
\pacs {
89.75.Fb, 
05.65.+b, 
89.65.Ef, 
89.75.Hc 
}
\maketitle

\section {1. Introduction}

      The naming game (NG) \cite{Baronc,Baronc4} is a simple multi-agent model that employs mutual bipartite interactions within a
   population of individuals which leads to the emergence of a shared communication scheme. In each game a randomly selected pair of agents 
   interact to negotiate conventions, i.e., associations between forms (names) and meanings (for example objects in the 
   environment, linguistic categories etc.). The negotiation of conventions takes place as follows: one of the agents 
   (acting as a speaker) attempts to draw attention of the other agent (acting as the hearer) toward the external meaning 
   (i.e., an object or a category) by the production of a conventional form. In case the hearer is able to express the 
   proper meaning of the form uttered by the speaker the agent pair is assumed to meet a mutual consensus and the interaction 
   is called a ``success". Consequently, both the agents update their form-meaning repertoire by removing all competing 
   forms corresponding to the meaning except the ``winning" one currently uttered by the speaker. On the other hand, if 
   the hearer produces a wrong interpretation then she takes lesson from the meeting by learning this new form-meaning 
   association and in this case the interaction is termed as a ``failure''. Thus, on the basis of success and failure of 
   the hearer in producing meaning of the name, both the interacting agents reshape their internal form-meaning association. 
   Through successive interactions, the adjustment of such individual associations collectively leads or should 
   lead to the emergence of a global consensus.

      The naming game model is one of the simplest examples of a framework progressively leading to the establishment of 
   human-like languages. It was initially formulated to understand the role of self-organization in the evolution and 
   change of human languages \cite{Steels1,Steels2}. Since then, this model has acquired a paradigmatic role in the novel field 
   of semiotic dynamics (see~\cite{Baronc,Baronc4,Baronc1,Baronc2,Baronc3,Dal,Luca} for a series of references) that primarily investigates 
   how language evolves through the invention and successive adoption 
   of new words and grammatical constructions. NG finds wide applications in various fields ranging from artificial sensor 
   network as a leader election model \cite{Baronc3} to the social media as an opinion formation model \cite{Maity}. More advanced
   models~\cite{Puglisi,Animesh} attempting to explain further complex processes like categorization have also been built on top of the 
   basic naming game framework.  

\begin{figure}[top]
\begin {center}
\includegraphics[width=7.0cm]{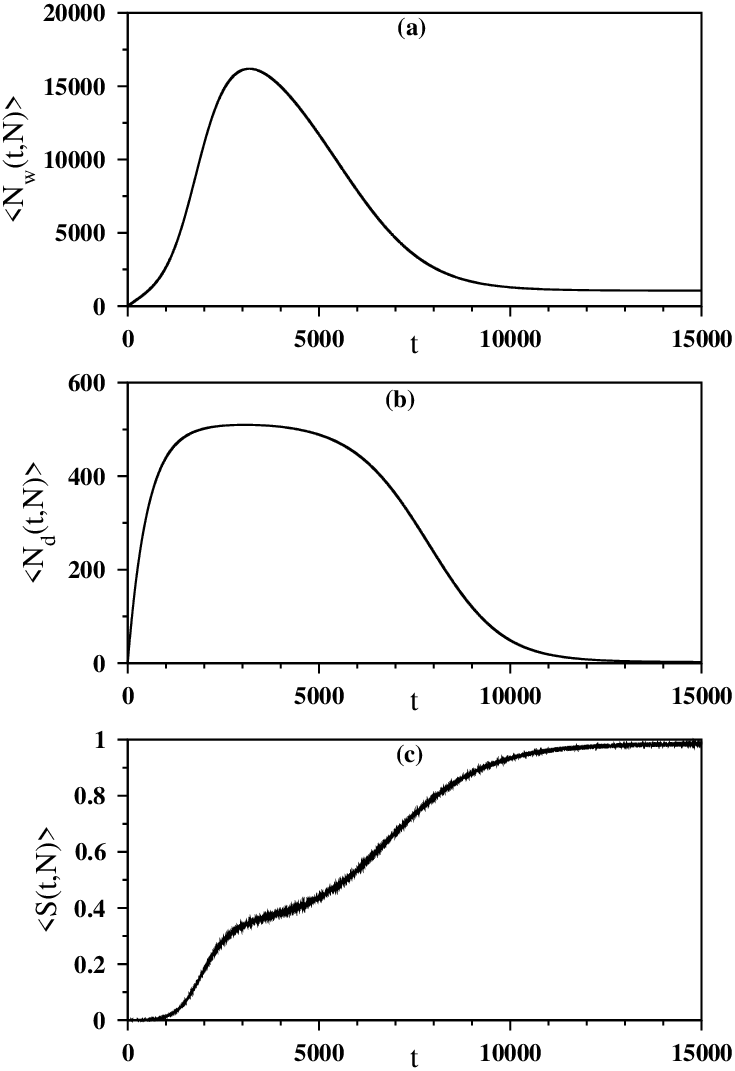}
\end {center}
\caption{Variations of the average
(a) number of words $\langle N_{w}(t,N) \rangle$,
(b) number of distinct words $\langle N_{d}(t,N) \rangle$ and
(c) success rate $\langle S(t,N) \rangle$
with time for a community size $N = 1024$.
}
\end{figure}

      In this paper, we revisit the basic construction of this model and argue that it is too stringent in removing all 
   the entries except the winning one from the agent repertoire after a successful interaction. Further, it has to be noted 
   that learning is seldom unidirectional as in case of a failure in the original naming game; in contrast, we believe that 
   learning activity most of the times is reciprocal \cite{Fitch,Hendriks}. Therefore, here we redefine the interaction 
   rules in order to address the above limitations by having a symmetric model where on a success both the agents sort out 
   all the common information that they have while on a failure enhance each of their knowledge by learning all the 
   form-meaning associations that the other partner only knew so far. One can intuitively posit that this process should lead 
   to the emergence of a faster consensus than the original naming game owing to the fact that (a) the agents learn more and 
   (b) the agreement criteria is relaxed, thereby, increasing manifolds the probability of successful communication. We 
   perform rigorous numerical simulations to obtain the scaling relations for this revised model and explicitly show that for 
   a population of $N$ agents the time to reach global consensus indeed scales as $N^{1.13}$ as opposed to $N^{1.5}$ for 
   the original naming game.  

\begin{figure}[top]
\begin {center}
\includegraphics[width=7.0cm]{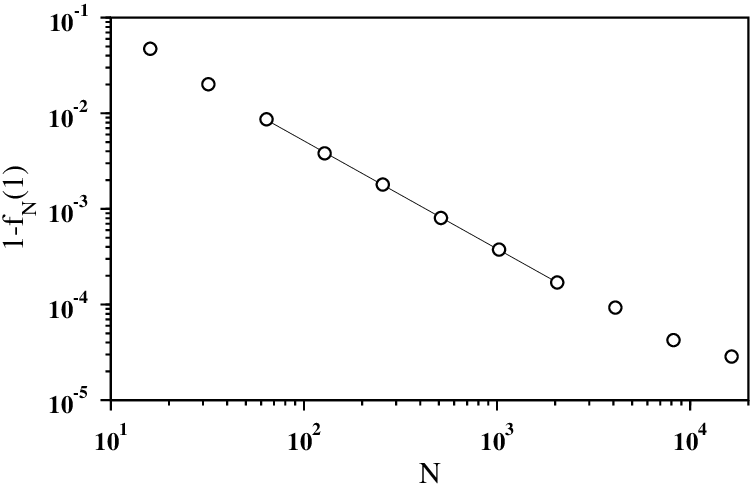}
\end {center}
\caption{
The fraction $1 - f_N(1)$ of configurations having more than 1 distinct word per agent
in the stable state has been plotted against the community size $N$. A power law is observed
like $1-f_N(1) \sim N^{-\tau}$ with $\tau \approx 1.13(2)$.
}
\end{figure}

\section {2. The Model}

      There are $N$ agents in a community. Each agent $i (i=1, ... , N)$ has an inventory of words whose length $\ell$ may be 
   arbitrarily long. The community evolves under a dynamical process where a long sequence of mutual bipartite information
   sharings takes place which finally reaches a stable state where all agents have the same set of common words. At the initial
   stage all agents have empty inventories i.e., $\ell_i=0$ for all $i$. In an interaction a pair of distinct 
   agents $i$ and $j$ of inventory lengths $\ell_i$ and $\ell_j$ respectively are selected randomly with uniform probability
   from all the agents. One of them, say the $i$-th agent is randomly selected between the two and is termed as the
   `speaker' where as the $j$-th agent is called the `hearer'. The time $t$ is discrete and is measured in terms of the number of
   interactions. The interaction between them can take place in the following three possible ways:

   A. {\it Invention:} In this case the inventory list of the speaker is empty. The speaker picks up a new 
   word and keeps it at the bottom of his inventory. Since this is a new word, it cannot be present in the inventory of 
   the $j$-th agent. Therefore this new word is simply added at the top of the inventory of the $j$-th agent.

\begin{figure}[top]
\begin {center}
\includegraphics[width=7.0cm]{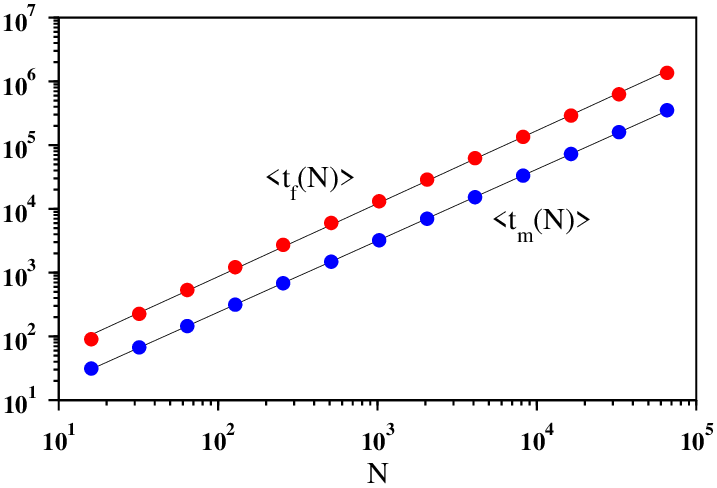}
\end {center}
\caption{(Color online) The variations of the average maximum time $\langle t_m(N) \rangle$ (blue) and the average 
convergence time $\langle t_f(N) \rangle$ (red) against the community size $N$. The exponents are $\alpha = 1.12$ and 
$\beta = 1.14$ respectively.
}
\end{figure}

   B. {\it Success:} In this case the inventory length of the speaker is non-zero. The speaker and the hearer share
   information about their contents, sort out the common contents and only the common words are retained. That means the inventories 
   of lengths $\ell_i$ and $\ell_j$ of the speaker and the hearer respectively are compared and the number $n$ of common 
   words are sorted out. In case $n>0$ then this possibility is called a success. The inventories of both the speaker
   and the hearer are then shrunk to $n$ entries where only the common words are kept.

   C. {\it Failure:} If the inventories have non-zero lengths yet there is no common word between them, then the 
   lists are merged together and both the agents have the same combined list.

      It is to be noted that in this model the success and failure rules are symmetric with respect to the speaker and the hearer.

      At any arbitrary intermediate time $t$ the total number of words in the community is denoted by $N_w(t)$ and the total 
   number of distinct
   words is denoted by $N_d(t)$. The dynamics starts with the inventory lengths $\ell_i=0$ for all $i$. At very early times 
   almost all interactions are of type A. During this period both $N_w(t)$ and $N_d(t)$ grow very fast, i.e., linearly with time. 
   As time proceeds more and more agents have non-zero inventory lengths and therefore the chances of interactions of types B and C 
   become increasingly likely. Consequently $N_w(t)$ reaches a maximum at a specific time $t_m$ and then it decreases with 
   time (Fig. 1(a)). On the other hand the number of distinct words $N_d(t)$ nearly saturates around a fixed mean value. Eventually
   $N_d(t)$ also decreases gradually and the community finally converges at the time $t_f$ to a stable state which is a fixed point 
   (Fig. 1(b)). In this stable
   state $N_w(t_f)$ takes a value $gN$ when every inventory has the same set of $g$ common words, $g$ being 
   a small positive integer. Therefore in contrast to the naming game model where $g$ =1 \cite {Baronc}, there could be multiple
   globally common words in our model i.e., $g > 1$. Consequently $N_d(t)$ finally reaches the value $g$. In addition a third quantity 
   ${\cal S}(t)$ is also calculated which measures the success rate of an interaction at time $t$. In other words ${\cal S}(t)$ 
   is the fraction of a large number of independent runs which have successful moves at time $t$ and its variation with time is 
   shown in Fig. 1(c).

\begin{figure}[top]
\begin {center}
\includegraphics[width=7.0cm]{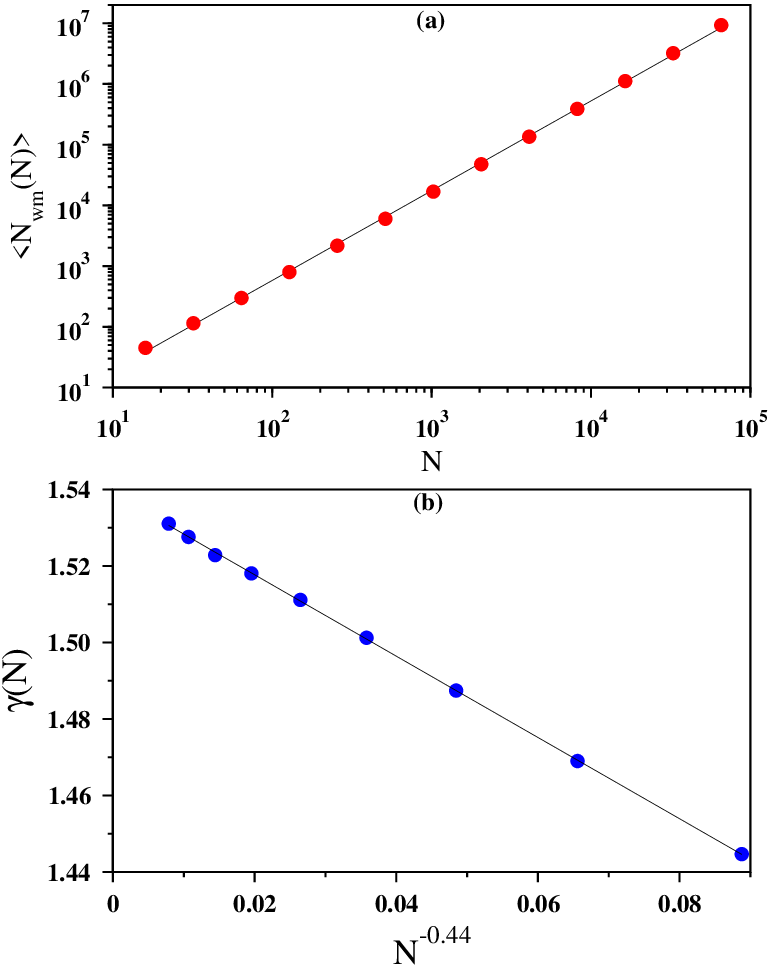}
\end {center}
\caption{(Color online) (a) Plot of the average maximal number of words $\langle N_{wm}(N) \rangle$ against the community size 
$N$ on a $\log - \log$ scale. (b) The slopes $\gamma(N)$ between pairs of successive points in (a) gradually increases with 
increasing $N$ and has been plotted against $N^{-0.44}$ to obtain the asymptotic value of $\gamma$ = 1.539.
}
\end{figure}

\section {3. The Algorithm}

      The simulation algorithm can be described as follows. Positive integer numbers starting from unity are used for representing
   different words. Therefore at any arbitrary intermediate stage if $N_d$ distinct words have already been used, to choose 
   a new word one simply selects the number $N_d+1$. It turned out that defining an array $b(k)$ is very useful, $b(k)$ keeps 
   track of the number of times the word $k$ has occurred with all agents. In case A, $b(k)$ is increased by 2: $b(k) \to b(k)+2$. 
   However to check if an interaction is a case of success or failure, one first compares the inventories of the $i$-th and the
   $j$-th agents. Therefore every word of the list $\ell_i$ has to be checked in the list $\ell_j$ and vice versa. This is 
   easily done by using another array $a(k)$ and for every word $k$ in $\ell_i$ and $\ell_j$ one makes $a(k) \to a(k)+1$.
   After that the number of locations with $a(k)=2$ are the number of common words between $\ell_i$ and $\ell_j$. Let this
   number be $n$ and only these common words are kept in another array $a_1$. At the same time we also count that out of $n$ such
   common words how many have $b$ values greater than 2, i.e., these words have not only occurred in $\ell_i$ and $\ell_j$ but
   also in the inventories of other agents. Let this number be $n'$. If $n > 0$ it is a case of success and if $n = 0$ its is 
   a case of failure. 

\begin{figure}[top]
\begin {center}
\includegraphics[width=7.0cm]{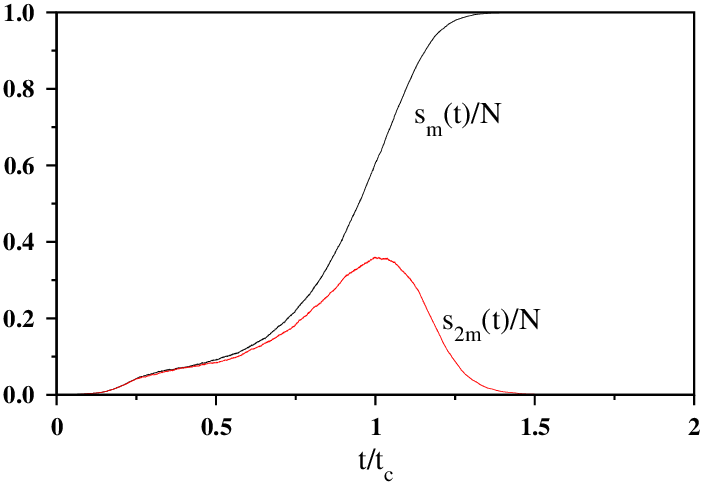}
\end {center}
\caption{
(Color online) For a single run the variations of the scaled sizes of the largest cluster $s_m(t)/N$ and the second largest cluster 
$s_{2m}(t)/N$ for a community with $N$ = 16384 agents. It is seen that while the size of the largest cluster grows almost (but not exactly)
monotonically, the size $s_{2m}(t)$ of the second largest cluster reaches a maximum at time $t_c$ and then gradually decreases to zero. 
The time axes has been scaled by the characteristic time $t_c$.
}
\end{figure}

      In case of success we first update the $b$ array. For each entry $k$ in $\ell_i$ and $\ell_j$ we first make 
   $b(k) \to b(k)-1$. Therefore during this updating procedure whenever $b(k)$ becomes zero we reduce $N_d$ by one: 
   $N_d \to N_d -1$. Let there be $m$ distinct entries in the inventories of $i$ and $j$ where $b(k)$ becomes zero.
   Then the $n$ words in $a_1$ array are copied to $\ell_i$ and $\ell_j$. $N_w$ is updated like: 
   $N_w \rightarrow N_w-\ell_i-\ell_j+2n$ and $N_d$ is updated like: $N_d \rightarrow N_d - m + n - n'$. This completes 
   a successful interaction.

      In the case of failure the combined list of $\ell_i$ and $\ell_j$ are copied to the inventory lists of $i$ and $j$. 
   For each such word the $b$ value is increased by unity. The total number of words $N_w$ is increased as $N_w \rightarrow
   N_w+\ell_i+\ell_j$, the number of distinct words $N_d$ remains same. This completes an unsuccessful interaction (see Table I).

\begin{table}[b]
\begin {tabular}{lll}
Rule A: & \hspace*{0.1cm} $N_w \rightarrow N_w+2$;                & \hspace*{0.1cm} $N_d \rightarrow N_d+1$ \\
Rule B: & \hspace*{0.1cm} $N_w \rightarrow N_w-\ell_i-\ell_j+2n$; & \hspace*{0.1cm} $N_d \rightarrow N_d - m +n-n'$ \\
Rule C: & \hspace*{0.1cm} $N_w \rightarrow N_w+\ell_i+\ell_j$;    & \hspace*{0.1cm} $N_d \rightarrow N_d$  \\ 
\end {tabular}
\caption{Summary of the rules A, B and C for the changes in the total number of words $N_w(t)$ and the number of 
distinct words $N_d(t)$ at any arbitrary time $t$.}
\end {table}

\section {4. The Results}

      It is noticed that on increasing the community size $N$ the probability that an arbitrary configuration has the same set of
   $g$ distinct words per agent in the final stable state decreases for $g > 1$ and it increases to unity for $g = 1$. We have
   measured the fraction $f_N(g)$ of a large sample of uncorrelated configurations that have $g$ words in the final stable
   configurations. The variation of $f_N(1)$ has been shown in Fig. 2. A plot of $1-f_N(1)$ vs. $N$ on a $\log - \log$ scale
   gives a nice straight line for the intermediate range of $N$. This indicates that the growth of $f_N(1)$ to unity as $N$
   increases follows a power law like $1-f_N(1) = AN^{-\tau}$ and our measured value of $\tau$ is 1.13(2). 

\begin{figure}[t]
\begin {center}
\includegraphics[width=7.0cm]{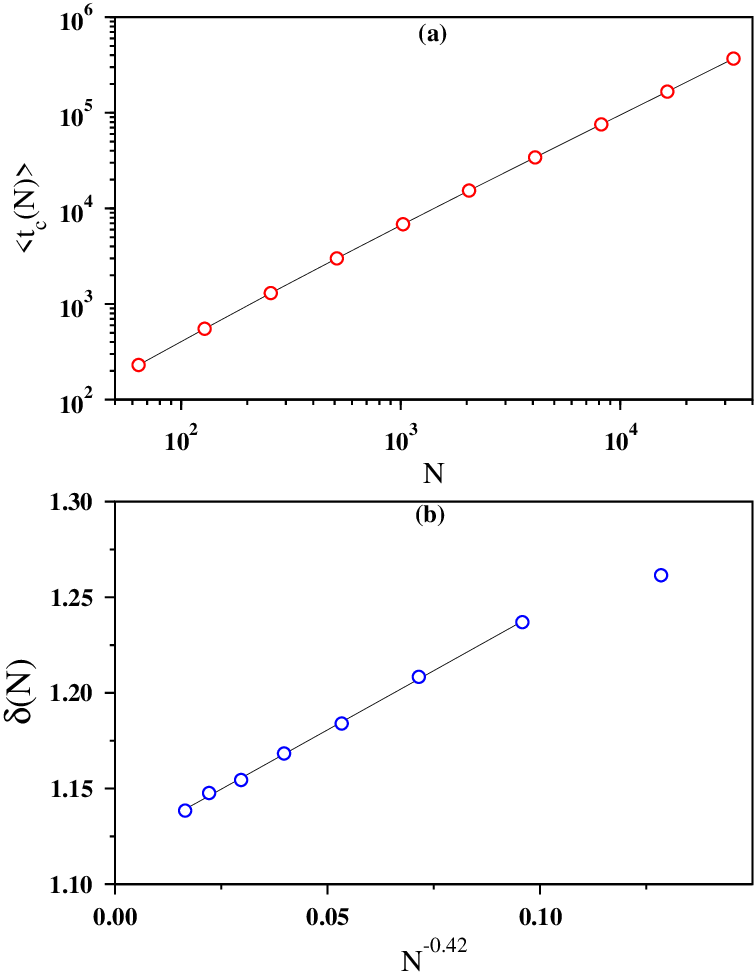}
\end {center}
\caption{
(Color online) 
(a) The average value of the characteristic time $\langle t_c(N) \rangle$ where the size of the second largest cluster is maximum
has been plotted with the community size $N$ on a $\log - \log$ scale for $N = 2^6, ..., 2^{15}$. The variation seems to be a
power law: $\langle t_c(N) \rangle \sim N^{\delta}$.
(b) Slopes between successive points has $N$ dependence and we plot $\delta(N)$ vs. $N^{-0.42}$ which fits best to a straight line.
The extrapolated value $\delta$ = 1.12.
}
\end{figure}

      The mean maximal time $\langle t_m(N) \rangle$ and the mean convergence time $\langle t_f(N) \rangle$ have been 
   measured for different values of $N$ and are plotted using a $\log - \log$ scale in Fig. 3. The community sizes which 
   have been simulated varied from $N = 2^4, 2^5, ...., 2^{16}$, increased by a factor of 2 in successive steps. These 
   data fit very well to straight lines. Therefore assuming power law variations like
\begin {equation}
\langle t_m(N) \rangle \sim N^{\alpha} \hspace*{0.7cm} \text{and} \hspace*{0.7cm}  \langle t_f(N) \rangle \sim N^{\beta}
\end {equation}
   we obtained $\alpha = 1.12$ and $\beta = 1.14$. 

      This observation leads us to conclude that both $\alpha$ and $\beta$ are approximately the same and has a value 1.13(1). 
   It may be noted that these exponents are much smaller than the original naming game (both $\alpha$ and $\beta$ equal to 1.5) 
   \cite {Baronc}. This faster consensus is possibly a consequence of the fact that the interaction rule here is symmetric thus
   increasing the possibility of alignment between the agents through fewer interactions as compared to the original naming game.
   Further, here the stable state criteria is also relaxed, so the agents are assumed to reach consensus even if they do not
   agree on only a single word.

\begin{figure}[top]
\begin {center}
\includegraphics[width=7.0cm]{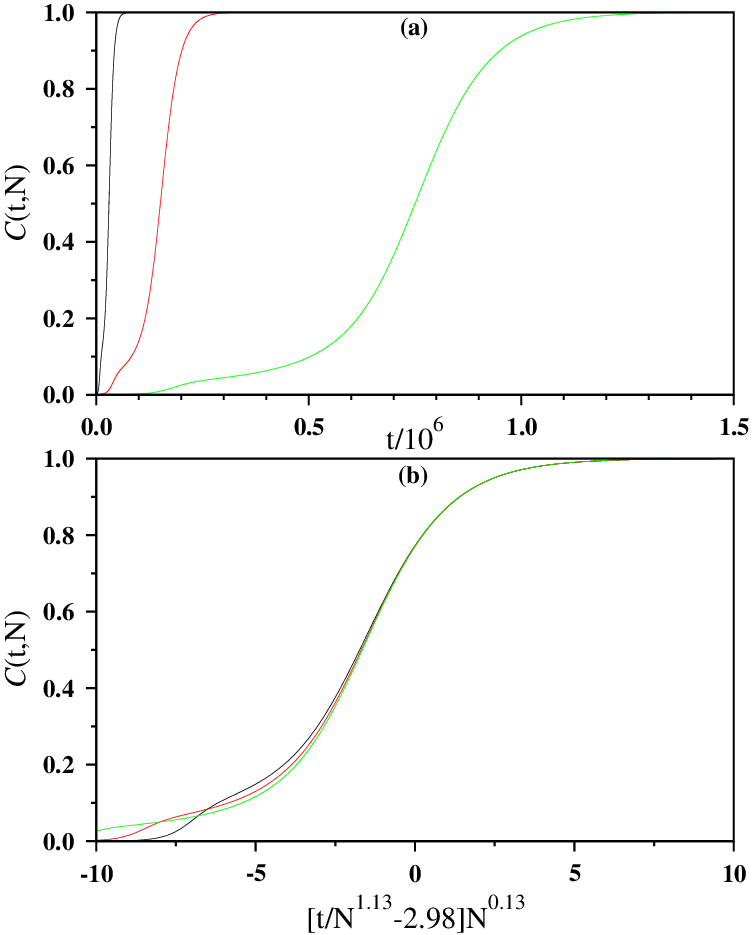}
\end {center}
\caption{
(Color online) 
(a) Variation of the fractional size of the largest cluster ${\cal C}(t,N) = \langle s_m(t,N) \rangle / N$ with time $t/10^6$.
(b) Finite size scaling of the data in (a), plot of ${\cal C}(t,N)$ vs. $[t/N^{1.13}-2.98]N^{0.13}$
exhibits a data collapse. 
}
\end{figure}

      Next in Fig 4. we plotted the average maximal number of words $\langle N_{wm}(N) \rangle$ against $N$ on a $\log - \log$ 
   scale for the same community sizes. Here again we assumed a power law variation like 
\begin {equation}
\langle N_{wm}(N) \rangle \sim N^{\gamma}
\end {equation}
   and the average slope 
   is measured using a least square fit method. We obtained an average value of $\gamma = 1.49$. Further, this analysis has been done
   in more detail. The intermediate slopes $\gamma(N)$ between successive pairs of points have been measured and extrapolated
   against $N^{-0.44}$. The extrapolation fits very well to a straight line and in the limit of $N \to \infty$ the
   value of $\gamma = \gamma(\infty) = 1.539$ has been obtained. This value of $\gamma$ is comparable with 1.5 in the
   original naming game model \cite {Baronc}.

      It may be noticed that the main difference between the present model and the original naming game arises from the fact that here a 
   failure is caused when none of the words known by the speaker is also known by the hearer as opposed to the mean field case 
   where the sufficient condition for failure is that the one random word selected by the speaker is unknown to the hearer. 
   A similar argument also holds for the success case where the success probability in this case is determined by whether the hearer 
   knows one or two or three or up to any number of words known by the speaker unlike the mean field case where the success is 
   determined by the match of the one random word selected by the speaker. Therefore, intuitively a single success or failure 
   event in our model corresponds to an accumulation of a set of a number of independent success and failure events in the mean 
   field case thus making the current dynamics faster and the overall exponents different.

\section {5. The Largest Cluster}

      At an intermediate time $t$ there are $N_d(t)$ distinct words and in general each word is shared by a number of agents. Similar to the 
   percolation phenomena \cite {Stauffer} we define the cluster size $s_i$ associated with the $i$-th word as the number of distinct agents which 
   have the word $i$ in their inventories. In the algorithm described in section 3 we have stored the cluster sizes in the array 
   $b(i)$. As time evolves cluster sizes of some words gradually vanish but at the same time the cluster sizes of the other words 
   grow. Finally only $g$ distinct words survive whose cluster sizes are exactly $N$ and at this point of time the dynamics reaches 
   the fixed point. It may be noted that the size of a particular cluster increases in the failure rule and decreases in the 
   success rule only by one agent at a time. We keep track of the variation of the size of the largest cluster $s_{m}(t,N)$ and observe 
   how it almost monotonically increases and assumes the size $N$ at the fixed point (Fig. 5).  At an intermediate stage there may 
   be a number of distinct clusters whose sizes are equal to the largest cluster size $s_m(t,N)$. We define the fractional size of 
   the largest cluster at time $t$ averaged over many independent runs as:
\begin {equation}
{\cal C}(t,N) = \langle s_{m}(t,N) \rangle /N.
\end {equation} 
   In addition we define the size $s_{2m}$ of the second largest cluster as well. In contrast to $s_m$ the value of $s_{2m}$ gradually
   increases to a maximum value and then systematically decreases to zero at the fixed point (Fig. 5). We define another characteristic 
   time $t_c$ at which the second largest cluster assumes its maximum value. This is the transition time when the second largest
   cluster starts dismantling and the largest cluster grows at its fastest rate which signifies the onset of correlation in the community. 

      In Fig. 6(a) the characteristic time $\langle t_c(N) \rangle$ averaged over many independent runs has been plotted on a 
   $\log - \log$ scale against the community sizes $N = 2^6, 2^7, ... , 2^{15}$. While the points seem to fit a nice straight 
   line on the average, a closer look reveals that here again the local slopes between successive pairs of points have a systematic 
   variation. Assuming that the functional form would indeed be a power law in the limit of $N \to \infty$ as 
\begin {equation}
\langle t_c(N) \rangle \sim N^{\delta}
\end {equation}
   we have extrapolated the local slopes $\delta(N)$ with a negative power of $N$. 
   The best value of this correction exponent is 0.42 and in Fig. 6(b) a plot of $\delta(N)$ against $N^{-0.42}$ gives a 
   nice straight line for large $N$ values. Extrapolating to $N \to \infty$ we obtained $\delta = 1.12$.

      Finally in Fig. 7 the average value of the largest cluster size ${\cal C}(t,N)$ has been plotted for three different 
   community sizes $N = 2^{12}, 2^{14}$ and $2^{16}$. We first scale the time axis $t/N^{1.13}$ so that the scaled time
   could be treated similar to the site / bond occupation probability in percolation theory. The scaled axis is then shifted
   by 2.98 and then again scaled by $N^{0.13}$ to obtain a data collapse.

      To summarize we devised a new model for information sharing and sorting in a community of agents. Three types of mutual bipartite
   interactions take place among the randomly selected pairs of agents. Here the interactions are more symmetric and less restricted compared 
   to the ordinary naming game. By Invention new words are created, by Failure inventories are shared and by Success only the common 
   words are sorted out. The dynamics of the system is dominated initially by Invention, followed by rapid growth of different words dominated 
   by Failure and finally the system gradually gets rid of uncommon words dominated by Success moves. The system finally reaches the stable 
   state where each agent has the same set of $g$ words in his inventory. Using extensive numerical studies we find that the exponents
   describing the characteristic time scales and the maximum number of words of this model assume a completely distinct set of values compared to 
   the ordinary naming game.

   E-mail: manna@bose.res.in


\begin{thebibliography}{90}
\bibitem {Baronc} A. Baronchelli, M. Felici, V. Loreto, E. Caglioti and L. Steels, J. Stat. Mech. (2006) P06014.
\bibitem {Baronc4} A. Baronchelli, V. Loreto, L. Steels, Int. J. Mod. Phys. C {\bf 19}, 785 (2008).
\bibitem {Steels1} L. Steels, Artificial Life, {\bf 2}, 318 (1995).
\bibitem {Steels2} L. Steels, p179 in {\it Artificial Life V }, 1996, Nara, Japan.
\bibitem {Baronc1} A. Baronchelli, L. Dall'Asta, A. Barrat and V. Loreto, Phys. Rev. E {\bf 73}, 015102R (2006).
\bibitem {Baronc2} L. Dall'Asta, A. Baronchelli, A. Barrat and V. Loreto, Phys. Rev. E {\bf 74}, 036105 (2006).
\bibitem {Baronc3} A. Baronchelli, Phys. Rev. E {\bf 83}, 046103 (2011).
\bibitem {Dal} L. Dall'Asta, A. Baronchelli, A. Barrat and V. Loreto, Europhys. Lett., {\bf 73}, 969 (2006).
\bibitem {Luca} L. Dall'Asta and A. Baronchelli, J. Phys. A, {\bf 39}, 14851 (2006).
\bibitem {Maity} S. K. Maity, T. V. Manoj and A. Mukherjee, Phys. Rev. E {\bf 86}, 036110 (2012).
\bibitem {Puglisi} A. Puglisi, A. Baronchelli and V. Loreto, Proc. Natl. Acad. Sci. {\bf 105}, 7936 (2008).
\bibitem {Animesh} V. Loreto, A. Mukherjee and F. Tria, Proc. Natl. Acad. Sci. 1113347109 (2012).
\bibitem {Fitch} W. T. Fitch, L. Huber and T. Bugnyar, Neuron {\bf 65}, 795 (2010).
\bibitem {Hendriks} P. Hendriks, H. de Hoop, I. Krämer, H. de Swart and Joost Zwarts, in {\it Conflicts in Interpretation} 2010, Equinox Publishing, London.
\bibitem {Stauffer} D. Stauffer and A. Aharony, {\it Introduction to Percolation Theory} (Taylor \& Francis, London, 1994).
\end {thebibliography}
\end {document}